\documentclass[superscriptaddress,floatfix]{revtex4}
\usepackage{color,graphicx,amsmath,hhline}
\usepackage[english]{babel}
\def\mc#1{\multicolumn{2}{c|}{#1}}
\def\mn#1{\multicolumn{2}{c}{#1}}

\def\Xv2{\ensuremath{{\dot X}^2}}
\def\Yv2{\ensuremath{{\dot Y}^2}}
\def\mtbar{\overline{m}_t}

\def\veta{\ensuremath{v_2(\eta)}\;}
\def\vpt{\ensuremath{v_2(p_t)}\;}
\def\e#1{\ensuremath{\times10^{-#1}}}

\begin{document}

\title{Universal scaling of the elliptic flow data at RHIC}
%\title{Universal scaling of the elliptic flow and the perfect hydro picture at RHIC}

\author{M. Csan\'ad}
\affiliation{Department of Atomic Physics, ELTE, Budapest,
P\'azm\'any P. 1/A, H-1117 Hungary}
\affiliation{Department
of Chemistry, SUNY Stony Brook, Stony Brook, NY, 11794-3400, USA}
\author{T. Cs\"org\H{o}}
\affiliation{MTA KFKI RMKI, H - 1525 Budapest 114, P.O.Box 49, Hungary}
\author{A. Ster}
\affiliation{MTA KFKI RMKI, H - 1525 Budapest 114, P.O.Box 49, Hungary}
\author{B. L\"orstad}
\affiliation{Department of Physics, University of Lund, S-22362 Lund, Sweden}
\author{N. N. Ajitanand}
\affiliation{Department of Chemistry, SUNY Stony Brook, Stony Brook, NY, 11794-3400, USA}
\author{J. M. Alexander}
\affiliation{Department of Chemistry, SUNY Stony Brook, Stony Brook, NY, 11794-3400, USA}
\author{P. Chung}
\affiliation{Department of Chemistry, SUNY Stony Brook, Stony Brook, NY, 11794-3400, USA}
\author{W. G. Holzmann}
\affiliation{Department of Chemistry, SUNY Stony Brook, Stony Brook, NY, 11794-3400, USA}
\author{M. Issah}
\affiliation{Department of Chemistry, SUNY Stony Brook, Stony Brook, NY, 11794-3400, USA}
\author{R. A. Lacey}
\affiliation{Department of Chemistry, SUNY Stony Brook, Stony Brook, NY, 11794-3400, USA}

\begin{abstract}
Recent PHOBOS measurements of the excitation function for the
pseudo-rapidity dependence of elliptic flow in Au+Au collisions at RHIC,
have posed a significant theoretical challenge. Here we show that these
differential measurements, as well as the RHIC measurements on transverse
momentum satisfy a universal scaling relation predicted by the Buda-Lund
model, based on exact solutions of perfect fluid hydrodynamics. We also
show that recently found transverse kinetic energy scaling of the elliptic
flow is a special case of this universal scaling.
\end{abstract}
\maketitle

\subsection{Introduction} One of the unexpected results from
experiments at the Relativistic Heavy Ion Collider (RHIC) is the
relatively strong second harmonic moment of the transverse momentum
distribution, referred to as the elliptic flow. Measurements of the
elliptic flow by the PHENIX, PHOBOS and STAR collaborations (see
refs.~\cite{Back:2004zg,Back:2004mh,Adler:2003kt,Adams:2004bi,Adler:2001nb,Sorensen:2003wi})
reveal rich details in terms of its dependence on particle type,
transverse ($p_t$) and longitudinal momentum ($\eta$) variables, and on
the centrality and the bombarding energy of the collision. In the soft
transverse momentum region ($p_t \lesssim 2$~GeV/c) measurements at
mid-rapidity are found to be well described by hydrodynamical
models~\cite{Adcox:2004mh,Adams:2005dq}. By contrast, differential
measurement of the pseudo-rapidity dependence of elliptic flow and its
excitation function have resisted several attempts at a description in
terms of hydrodynamical models (but see their description by the SPHERIO
model~\cite{Grassi:2005pm,Andrade:2007aa} or the approximate descriptions
in refs.~\cite{Nonaka:2007nn,Bleibel:2006xx}). Here we show that these
data are consistent with theoretical, analytic predictions that are based
on perfect fluid hydrodynamics: Fig.~\ref{f:v2w} demonstrates that the
investigated PHOBOS, PHENIX and STAR
data~\cite{Back:2004zg,Back:2004mh,Adler:2003kt,Adams:2004bi} follow the
theoretically predicted scaling law.

\subsection{Perfect fluid hydro picture} Perfect fluid
hydrodynamics is based on local conservation of entropy $\sigma$
and four-momentum tensor $T^{\nu\mu}$,
\begin{eqnarray}
    \partial_\mu (\sigma u^\mu) & = & 0, \\
    \partial_\nu T^{\mu \nu} & = & 0,
\end{eqnarray}
where $u^\mu$ stands for the four-velocity of the matter. The
fluid is perfect if the four-momentum tensor is diagonal in the
local rest frame,
\begin{equation}
    T^{\mu \nu}  =  (\epsilon + p) u^\mu u^\nu - p g^{\mu \nu}.
\end{equation}
Here $\epsilon$ stands for the local energy density and $p$ for the
pressure. These equations are closed by the equation of state, which gives
the relationship between $\epsilon$, $p$ and $\sigma$, typically $\epsilon
= \kappa p$ is assumed, where $\kappa$ is either a
constant~\cite{Csorgo:2003ry} or an arbitrary temperature dependent
function~\cite{Csorgo:2001xm} that uses a non-relativistic approximation.
Note also, that a bag constant can also be introduced, and the
$\epsilon-B=\kappa(p+B)$ equation of state can be
used~\cite{Csorgo:2003rt,Nagy:2007xn}.

We focus here on the analytic approach in exploring the consequences of
the presence of such perfect fluids in high energy heavy ion experiments
in Au+Au collisions at RHIC. Such a nonrelativistic exact analytic
solution was published in ref.~\cite{Csorgo:2001xm}, while relativistic
solutions were published in
refs.~\cite{Csorgo:2003rt,Sinyukov:2004am,Nagy:2007xn}. For a detailed
discussion on new exact relativistic solutions, see
ref.~\cite{Nagy:2007xn}.

A tool, that is based on the above listed exact, dynamical hydro
solutions, is the Buda-Lund hydro model of refs.
~\cite{Csorgo:1995bi,Csanad:2003qa}.  This hydro model is
successful in describing experimental data on single particle
spectra and two-particle correlations
~\cite{Csanad:2003sz,Csanad:2004cj}.  The model is defined with
the help of its emission function; to take into account the
effects of long-lived resonances, it utilizes the core-halo
model~\cite{Csorgo:1994in}.

The elliptic flow is an experimentally measurable observable and
is defined as the azimuthal anisotropy or second
fourier-coefficient of the one-particle momentum distribution
$N_1(p)$. The definition of the flow coefficients is:
\begin{equation}
v_n = \frac{\int_0^{2 \pi} N_1(p) \cos(n\varphi) d\varphi}
           {\int_0^{2 \pi} N_1(p) d\varphi},
\end{equation}
where $\varphi$ is the azimuthal angle of the momentum. This
formula returns the elliptic flow $v_2$ for $n=2$.

\subsection{Universal scaling of the elliptic flow in the Buda-Lund model}
The result for the elliptic flow, that comes directly from a
perfect hydro solution is the following simple scaling
law~\cite{Csorgo:2001xm,Csanad:2003qa}
\begin{equation}
v_2=\frac{I_1(w)}{I_0(w)},\label{e:v2w}
\end{equation}
where $I_n(z)$ stands for the modified Bessel function of the second kind,
$I_n(z) = (1/\pi)\int_0^\pi \exp(n \cos(\theta)) \cos(n \theta) d\theta$.

Note that this prediction was derived first in 2001 in a non-relativistic
perfect hydrodynamical solution, see eq.~(25) of
ref.~\cite{Csorgo:2001xm}. In 2003, it has been extended to the
relativistic kinematic domain in ref.~\cite{Csanad:2003qa}.
Ref.~\cite{Csanad:2003qa} considered a relativistic parameterization that
included a multitude of known relativistic solutions, such as the
Hwa-Bjorken solution~\cite{Hwa:1974gn,Bjorken:1982qr} or other
accelerating and Hubble-type of
solutions~\cite{Csorgo:2003rt,Sinyukov:2004am,Nagy:2007xn}, and
interpolated among these.

The subject of our current investigation is the testing of this scaling
law against recent experimental data, but let us first discuss this
scaling law.

In the Buda-Lund hydro model, elliptic flow depends only on momentum space
anisotropy, but it does not depend on the coordinate space anisotropy.
This feature of the Buda-Lund model~\cite{Csanad:2008af} is different from
azimuthally sensitive Blast-wave models~\cite{Retiere:2003kf}, where $v_2$
depends also on the coordinate space distribution.

In section~\ref{s:scvar}, we explain the universal scaling variable $w$ in
eq.~(\ref{e:v2w}) and show, how $w$ can be determined from measurements.
In section~\ref{s:exptest}, we shall subject this relationship to an
experimental test also.

\subsection{Variable of the universal scaling law of $v_2$}\label{s:scvar}
Looking at eq.~(\ref{e:v2w}), one sees that the Buda-Lund hydro model
predicts~\cite{Csorgo:2001xm} a \emph{universal scaling:} every $v_2$
measurement is predicted to fall on the same scaling curve $I_1/I_0$ when
plotted against the scaling variable $w$. This means, that $v_2$ depends
on any physical parameter (transverse or longitudinal momentum, mass,
center of mass energy, collision centrality, type of the colliding nucleus
etc.) only through the scaling variable $w$. This scaling variable is
defined by:
\begin{equation}
w = \frac{E_K}{T_*} \varepsilon \label{e:w}
\end{equation}
Here $E_K$ is a relativistic generalization of the transverse kinetic
energy, defined as
\begin{equation}
E_K = \frac{p_t^2}{2 \overline{m}_t},
\end{equation}
with
\begin{equation}
\overline{m}_t = m_t \cosh\left(\frac{y}{1+\Delta \eta\frac{m_t}{T_0}}\right),
\end{equation}
$y$ being the rapidity, $\Delta \eta$ the longitudinal expansion
of the source, $T_0$ the central temperature at the freeze-out and
$m_t = \sqrt{p_t^2+m^2}$ the transverse mass. We note, that at
mid-rapidity and for a leading order approximation, $E_K \approx
m_t - m$, which also explains recent development on scaling
properties of $v_2$ by the PHENIX experiment at
midrapidity~\cite{Adare:2006ti,Afanasiev:2007tv}. We furthermore
note, that parameter $\Delta\eta$ has recently been dynamically
related~\cite{Nagy:2007xn} to the acceleration parameter of new
exact solutions of relativistic hydrodynamics, where the
accelerationless limit corresponds to a Bjorken type, flat
rapidity distribution and the $\Delta\eta \rightarrow \infty$ limit.

The scaling variable $w$ also depends on the parameter $T_*$, which is the
effective, rapidity and transverse mass dependent slope of the azimuthally
averaged single particle spectra, and on the final momentum space
eccentricity parameter, $\varepsilon$. These can be
defined~\cite{Csorgo:2001xm,Csanad:2003qa} by the transverse mass and
rapidity dependent slope parameters of the single particle spectra in the
impact parameter (subscript $_x$) and out of the reaction plane (subscript
$_y$) directions, $T_x$ and $T_y$,
\begin{eqnarray}
    \frac{1}{T_*} &  = & \frac{1}{2}\left(\frac{1}{T_x}+\frac{1}{T_y}\right), \\
    \varepsilon & = & \frac{T_x - T_y}{T_x + T_y}.
\end{eqnarray}
which are thus observable quantities. Note also, that $\varepsilon$ can
also be interpreted as a measure of integrated $v_2$, and thus setting the
absolute scale of $v_2$.

\begin{figure}
\begin{center}
  \includegraphics[height=0.7\linewidth,angle=-90]{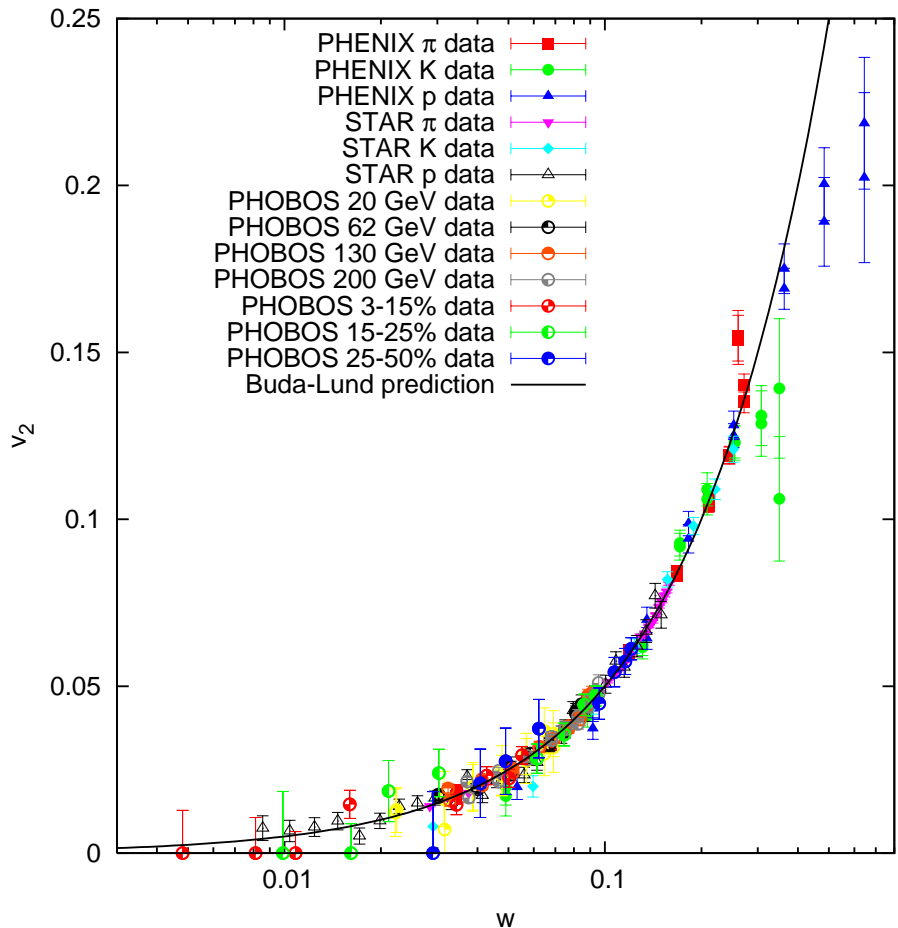}\\
\end{center}
\caption{Elliptic flow data of previous plots versus variable $w$ is shown:
Data points show the predicted~\cite{Csanad:2003qa} universal scaling.
Small scaling violations at large $w$ values correspond to \vpt data for
$p_t > 2 $ GeV. Note, that the error of $w$ was not plotted on this plot,
but it can be determined from data analysis, and it is on the order of 5-20\%.}
  \label{f:v2w}
\end{figure}

In the Buda-Lund hydro model~\cite{Csorgo:2001xm,Csanad:2003qa}, the
rapidity and the transverse mass dependence of the slope parameters is
given as
\begin{eqnarray}
    T_x & = & T_0 + \mtbar \dot X^2 \frac{T_0}{T_0 + \mtbar a^2}, \\
    T_y & = & T_0 + \mtbar \dot Y^2 \frac{T_0}{T_0 + \mtbar a^2}.
\end{eqnarray}
Here $a^2 = \langle \frac{\Delta T}{T}\rangle $ measures the
transverse temperature inhomogeneity of the particle emitting
source in the transverse direction at the mean freeze-out time.

We note, that each of the kinetic energy term, the effective
temperature $T_*$ and the eccentricity $\varepsilon$ are
transverse mass and rapidity dependent factors. However, for
$\overline{m}_t a^2 \gg T_0$, $T_x$ and $T_y$, hence $\varepsilon$
and $T_*$ become independent of transverse mass and rapidity. This
saturation of the slope parameters happens only if the temperature
is inhomogeneous, ie $a^2>0$.

The above structure of $w$, the variable of the universal scaling
function of elliptic flow suggests that the transverse momentum,
rapidity, particle type, centrality, colliding energy, and
colliding system dependence of the elliptic flow is only apparent
in perfect fluid hydrodynamics: a data collapsing behavior sets in
and a universal scaling curve emerges, which coincides with the
ratio of the first and zeroth order modified Bessel
functions~\cite{Csorgo:2001xm,Csanad:2003qa}, when $v_2$ is
plotted against the scaling variable $w$.

Interesting is furthermore, that the Buda-Lund hydro model also
predicts the following universal scaling laws and relationships
for higher order flows~\cite{Csanad:2003qa}: $v_{2n} = I_n(w)/I_0(w)$ and $v_{2n+1}
= 0$. This is to be tested in a later, more detailed analysis.

\subsection{Comparison to experimental data}\label{s:exptest}
We emphasize first, that the scaling variable $w$ is expressed in
eq.~(\ref{e:w}) in terms of factors, that are in principle measurable
(however, these factors are not yet determined directly from experimental
data). The elliptic flow $v_2$ is also directly measurable. Hence the
universal scaling prediction, eq.~(\ref{e:v2w}) can in principle be
subjected to a direct experimental test. Given the fact that such
measurements were not yet published in the literature, we perform an
indirect testing of the prediction, by determining the relevant parameters
of the scaling variable $w$ from an analysis of the transverse momentum
and rapidity dependence of the elliptic flow in Au+Au collisions at RHIC.

Transverse momentum dependent elliptic flow data at mid-rapidity can be
compared to the Buda-Lund universal scaling prediction of 2001 and 2003 of
the Buda-Lund model directly, as it was done in e.g.
ref.~\cite{Csanad:2003qa}.

Eq.~(\ref{e:v2w}) depends, for a given centrality class, on
rapidity $y$ and transverse mass $m_t$. When comparing our result
to \veta data of the PHOBOS Collaboration, we have performed a
saddle point integration in the transverse momentum variable and
performed a change of variables to the pseudo-rapidity $\eta=0.5
\log(\frac{|p| + p_z}{|p| - p_z})$, similarly to
ref.~\cite{Kharzeev:2001gp}. This way, we have evaluated the
single-particle invariant spectra in terms of the variables $\eta$
and $\phi$, and calculated \veta from this distribution, a
procedure corresponding to the PHOBOS measurement described in
ref.~\cite{Back:2004zg}.

Scaling implies data collapsing behavior, and also is reflected in a
difficulty in extracting the precise values of these parameters from
elliptic flow measurements: due to the data collapsing behavior, some
combinations of these fit parameters become relevant, other combinations
become irrelevant quantities, that cannot be determined from measurements.
This is illustrated in Fig. 1, where we compare the universal scaling law
of eq.~(\ref{e:v2w}) with elliptic flow measurements at RHIC. This figure
shows an excellent agreement between data and prediction. We may note the
small scaling violations at largest $w$ values, that correspond to
elliptic flow data taken in the transverse momentum region of $p_t > 2 $
GeV.

The observed scaling itself shows, that only a few relevant
combinations of $T_0$, $a^2$, $\dot X^2$, $\dot Y^2$ determine the
transverse momentum dependence of the $v_2$ measurements. Hence
from these measurements it is not possible to reconstruct all
these four source parameters uniquely. We have chosen the
following to eq.~(\ref{e:v2w}) approximative formulas to describe
the scaling of the elliptic flow:
\begin{eqnarray}
w(\eta) &=& \frac{2A}{\cosh(B\eta)}\textrm{, and}\label{e:we}\\
w(p_t) &=& A'\frac{p_t^2}{4m_t}\left(1+B'(m_t-m)+C'(m_t-m)^2\right),\label{e:wp}
\end{eqnarray}
and for small values of $w$ eq.~(\ref{e:v2w}) simplifies to $v_2
\approx w/2 $. The coefficients are as follows:
\begin{eqnarray}
A &=& \left.\frac{E_K}{2T_*}\varepsilon\right|_{m_t=\langle m_t \rangle,y=0}\label{e:A}\\
B &=& \left.\left(1+\Delta \eta \frac{m_t}{T_0}\right)^{-1}\right|_{m_t=\langle m_t \rangle,y=0}\\
A' &=& \left.\frac{2\varepsilon}{T_*}\right|_{m_t=m,y=0}\\
B' &=& \left.-\frac{1}{m}\frac{T_0}{T_0+ma^2}\left(1-2\frac{T_0}{T_*}\right)\right|_{m_t=m,y=0}\\
C' &=& \frac{1}{m}\left(\frac{T_0}{T_0+ma^2}\right)^5\left.\frac{1}{T_x^2T_y^2}\right|_{m_t=m,y=0}\times\label{e:C}\\
   &\times& \left[(\Xv2+a^2+\Yv2)(T_0+ma^2)^3+\right.\nonumber\\
   &+& m\Xv2\Yv2 \left(m^2(\Xv2\Yv2+a^2(\Xv2+\Yv2))\right)\nonumber\\
   &-&\left.3m\Xv2\Yv2T_0(T_0+ma^2)\right].\nonumber
\end{eqnarray}

From this simple picture we had to deviate a little bit in case of proton
\vpt data, here only one parameter could have been used to find a valid
Minuit~\cite{James:1975dr} minimum, so we fixed B' there.

For the case of kaons and protons, only $A'$ and $B'$ were significant,
while pion data were so detailed, that $C'$ could have been determined,
too. Thus we used it only when fitting pion \vpt.

For the analysis of the PHOBOS  \veta measurements at RHIC, we
have excluded points with large rapidity from lower center of mass
energies \veta fits ($\eta>4$ for 19.6\,GeV, $\eta>4.5$ for
62.4\,GeV). Points with large transverse momentum ($p_t>2.0$\,GeV)
were excluded from PHENIX and STAR \vpt fits. These values give a
hint at the boundaries of the validity of the model.

\begin{figure}
\begin{center}
  \includegraphics[height=0.7\linewidth,angle=-90]{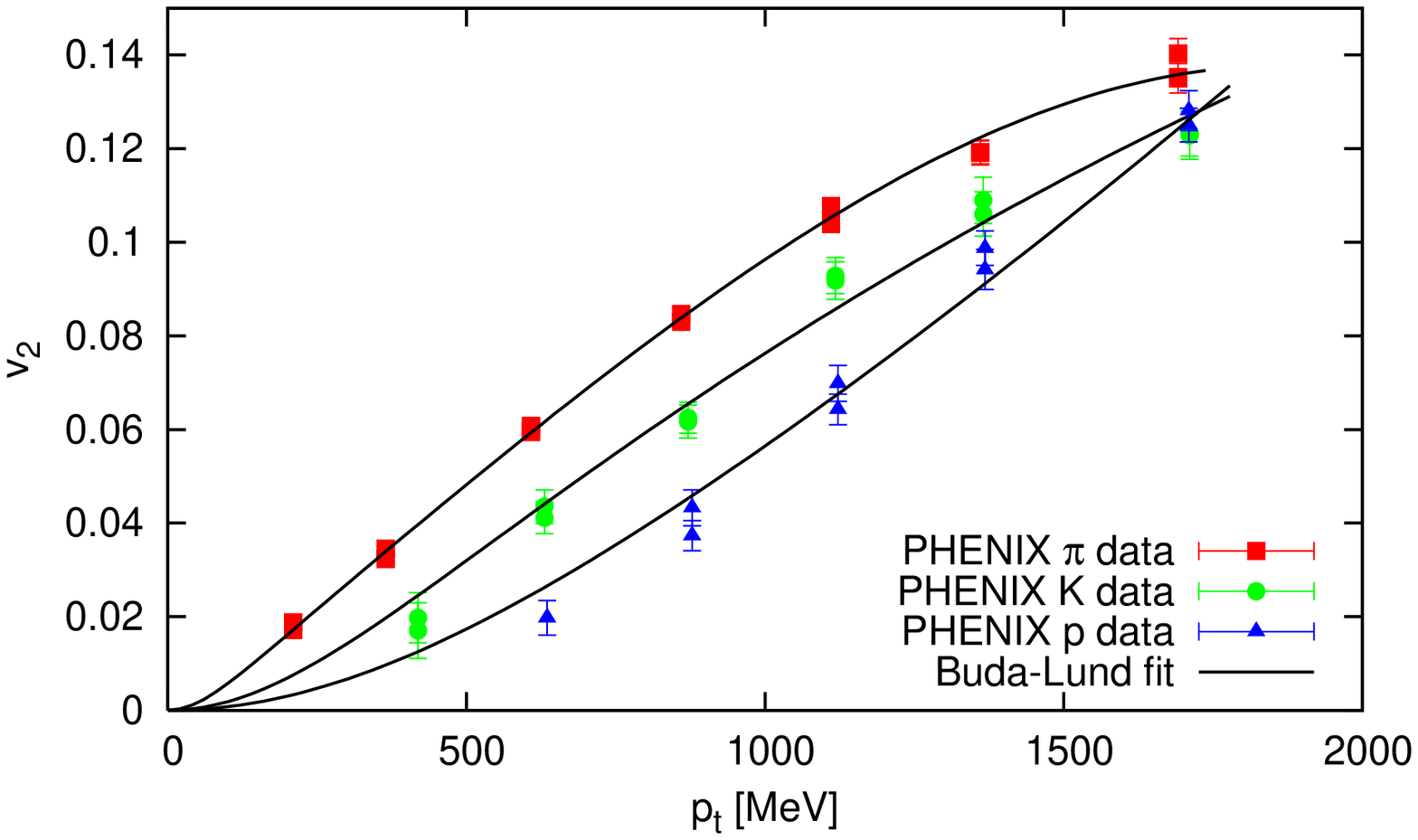}
  \includegraphics[height=0.7\linewidth,angle=-90]{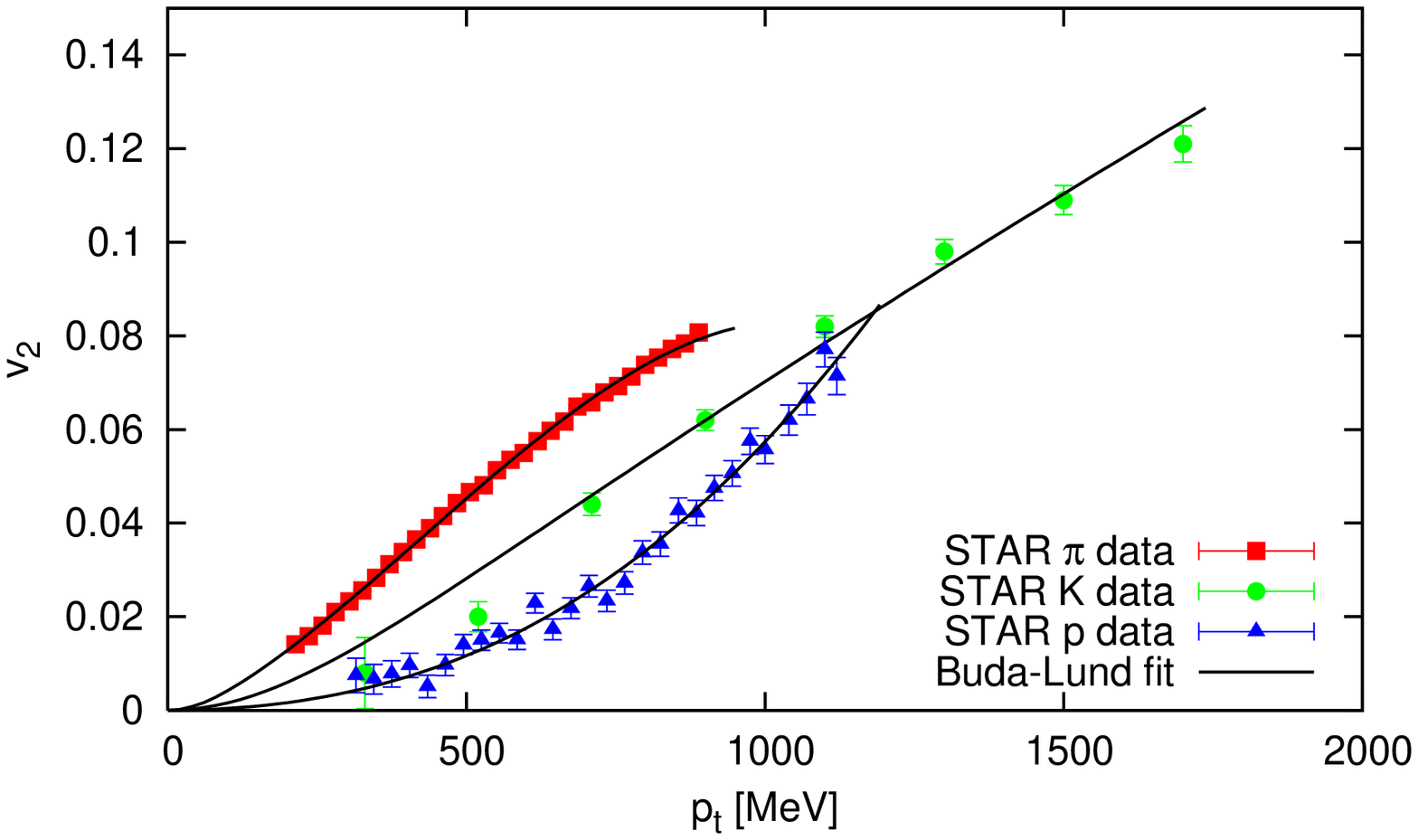}\\
\end{center}
  \caption{
PHENIX~\cite{Adler:2003kt} and
STAR~\cite{Adams:2004bi}
data on elliptic flow, $v_2$, plotted versus $p_t$ and fitted with Buda-Lund model.}
  \label{f:v2pt}
\end{figure}

\begin{figure}
\begin{center}
  \includegraphics[height=0.7\linewidth,angle=-90]{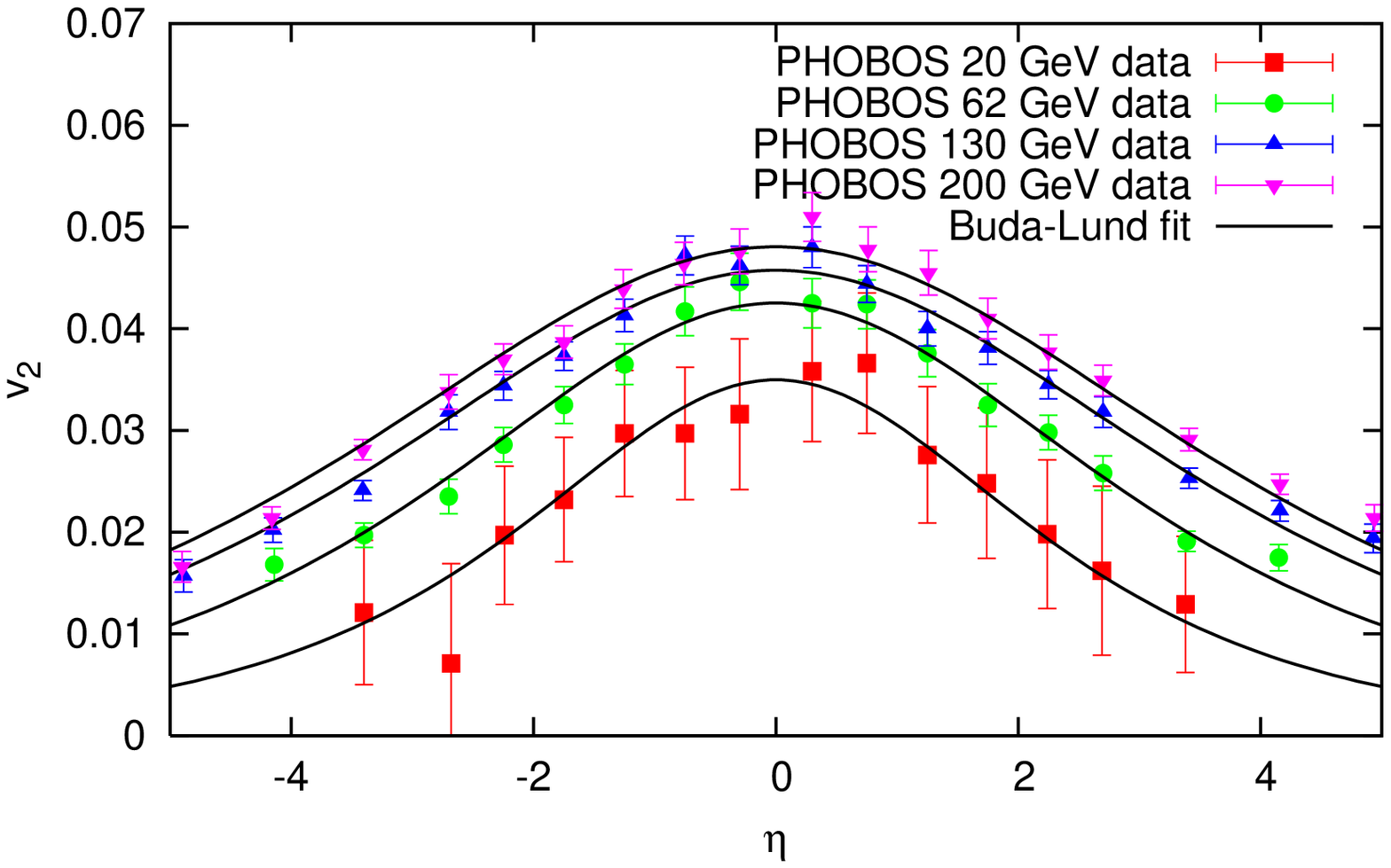}
  \includegraphics[height=0.7\linewidth,angle=-90]{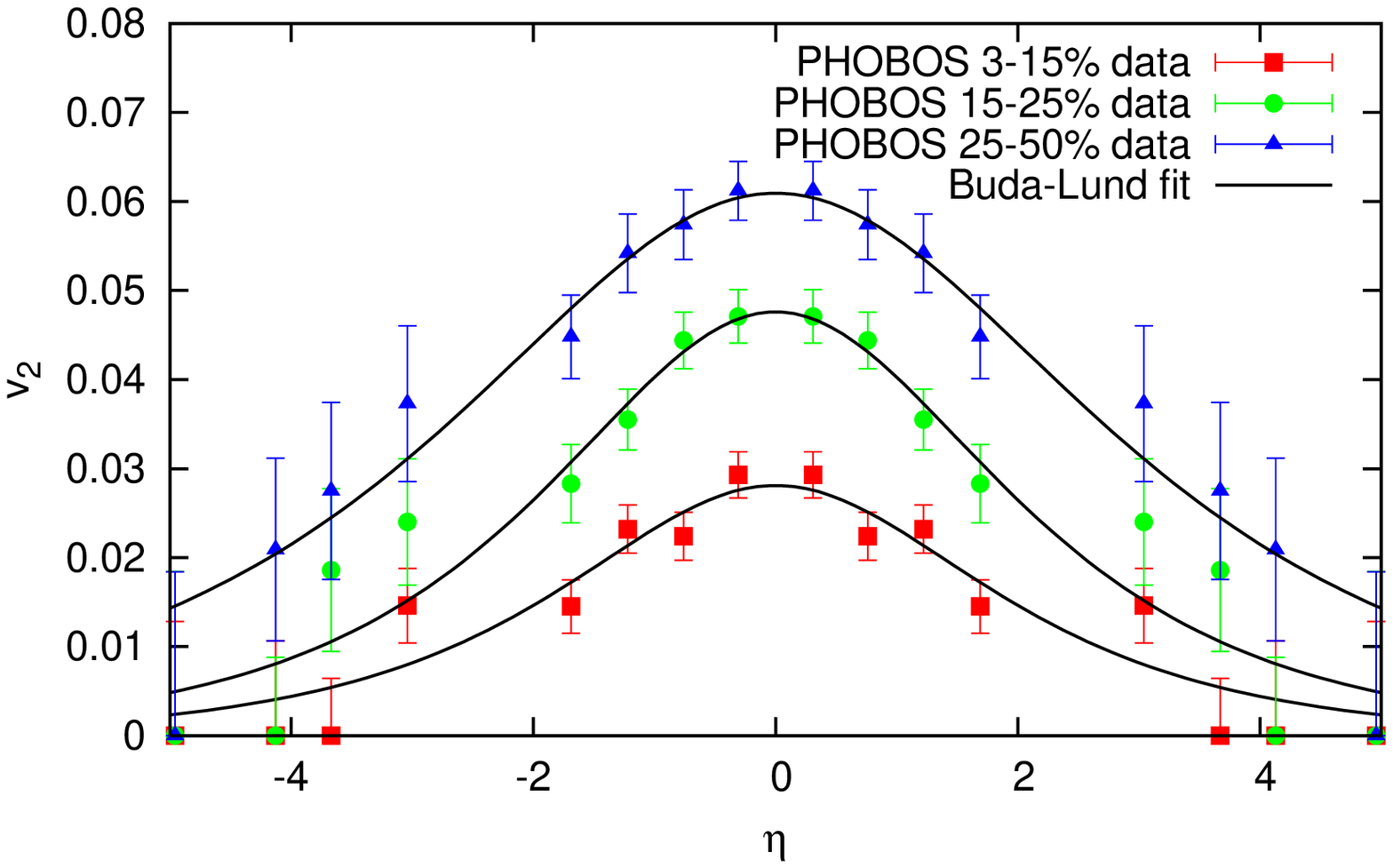}\\
\end{center}
  \caption{
PHOBOS~\cite{Back:2004zg,Back:2004mh} data on elliptic flow,
$v_2$, plotted versus $\eta$ and fitted with Buda-Lund model.}
  \label{f:v2eta}
\end{figure}

Fits to PHOBOS~\cite{Back:2004zg,Back:2004mh}, PHENIX~\cite{Adler:2003kt}
and STAR~\cite{Adams:2004bi} data are shown in
Figs.~\ref{f:v2pt}~and~\ref{f:v2eta}. The values of the parameters and the
quality of the fits are summarized in Table~\ref{t:vals}.

Using the fit parameters in Table~\ref{t:vals} and
eqs.~(\ref{e:we}-\ref{e:wp}), we have determined the universal scaling
variable $w$ from these PHENIX, PHOBOS and STAR data.

Using these $w$ values we have plotted the data against $w$ in
Fig.~\ref{f:v2w}. Note that on this plot $w$ itself has an error, as it is
a reconstructed variable based on eqs.~(\ref{e:A}-\ref{e:C}) and the
values and the errors of the parameters as given in Table~\ref{t:vals}.
Standard error propagation has been applied and we have obtained that the
relative error of $w$ changes between 5-50\%, and is above 20\% only for
\veta points with large $\eta$ and pion \vpt points with large $p_t$. As
Fig.~\ref{f:v2w} is best seen when the values of $w$ are plotted on a
logarithmic scale, these relative errors of $w$ do not change the
qualitative conclusion. Comparing with the solid black line in
Fig.~\ref{f:v2w} we observe, that elliptic flow data from various STAR,
PHENIX and PHOBOS measurements at RHIC follow the universal scaling curve,
predicted by the ellipsoidally symmetric Buda-Lund model in 2004.

\begin{table}
\begin{tabular}{|c|r@{$\pm$}l|r@{$\pm$}l|r@{$\pm$}l|r@{$\pm$}l|}\hline
  \veta               & \mc{20GeV}  & \mc{62GeV}  & \mc{130GeV}    & \mc{200GeV}  \\\hline
  A                   & 0.035& 0.004& 0.043& 0.001& 0.046 & 0.001  & 0.048& 0.001 \\\hline
  B                   & 0.53 & 0.1  & 0.41 & 0.01 & 0.34 &0.01     & 0.33 & 0.01  \\\hline
$\chi^2/$N$_{\rm DF}$ & \mc{1.7/11} & \mc{9.3/13} & \mc{17/15}     & \mc{18/15}   \\\hline
CL   & \mc{91\%}      & \mc{74\%}   & \mc{30\%}      & \mc{28\%}    \\\cline{1-9}\hhline{=======}
  \veta               & \mc{3-15\%} & \mc{15-25\%}& \mc{25-50\%}   & \mn{} \\\cline{1-7}
  A                   & 0.028& 0.002& 0.048& 0.002& 0.061 & 0.002  & \mn{} \\\cline{1-7}
  B                   & 0.64 & 0.08 & 0.60 & 0.06 & 0.43 & 0.04    & \mn{} \\\cline{1-7}
$\chi^2/$N$_{\rm DF}$ & \mc{12/13}  & \mc{8/13}   & \mc{4/13}      & \mn{} \\\cline{1-7}
CL                    & \mc{51\%}   & \mc{84\%}   & \mc{96\%}      & \mn{} \\\cline{1-7}
\end{tabular}
\vskip 2pt
\begin{tabular}{|c|r@{$\pm$}l|r@{$\pm$}l|r@{$\pm$}l|}\hline
  \vpt                & \mc{$\pi$}  & \mc{K}      & \mc{p}         \\\hline
A' [$10^{-4}$/MeV]    & 5.4  & 0.1  & 6.4  & 0.3  & 3.0  & 0.1     \\\hline
B' [$10^{-4}$/MeV]    & 16   & 1    &-0.2  & 0.4  & \mc{25, fixed} \\\hline
C' [$10^{-6}$/MeV$^2$]&-1.5  & 0.1  & \mc{---}    & \mc{---}       \\\hline
$\chi^2/$N$_{\rm DF}$ & \mc{96/27}  & \mc{17/5}   & \mc{27/26}     \\\hline
CL                    &\mc{1\e{7}\%}& \mc{0.5\%}  & \mc{40\%}      \\\hhline{=======}
  \vpt                & \mc{$\pi$}  & \mc{K}      & \mc{p}         \\\hline
A' [$10^{-4}$/MeV]    & 7.8 &0.2    & 7.4 & 0.3   & 5.8 & 0.1      \\\hline
B' [$10^{-4}$/MeV]    & 1.4 &0.6    &-1.3 & 0.4   & \mc{1.6, fixed}\\\hline
C' [$10^{-7}$/MeV$^2$]&-1.6 &0.3    & \mc{---}    & \mc{---}       \\\hline
$\chi^2/$N$_{\rm DF}$ & \mc{21/10}  & \mc{13/9}   & \mc{17/7}      \\\hline
CL                    & \mc{2\%}    & \mc{15\%}   & \mc{2\%}       \\\hline
\end{tabular}
  \centering \caption{Values of the parameters and the quality of the fits
  for collision energy dependent PHOBOS \veta data~\cite{Back:2004zg} is
  shown in the top table, the same for centrality dependent PHOBOS \veta
  data~\cite{Back:2004mh} in the second table. The third shows
  STAR~\cite{Adams:2004bi}, the fourth PHENIX~\cite{Adler:2003kt} \vpt
  data results.}\label{t:vals}
\end{table}

\section{Further scaling properties}
From the Taylor expansion of the Bessel functions for the realistic $w\ll
1$ case one finds
\begin{equation}
v_2\approx\frac{w}{2}.
\end{equation}
One can also see, that for small momenta,
\begin{equation}
E_K=\frac{p_t^2}{2 m_t}\approx m_t-m.
\end{equation}
Thus a leading order calculation, at mid-rapidity, from eq.~(\ref{e:wp})
one gets
\begin{equation}
v_2\approx\frac{A'}{4}(m_t-m).
\end{equation}
This derivation indicates, that the PHENIX
discovery~\cite{Adare:2006ti,Afanasiev:2007tv} of the universal scaling of
the elliptic flow in terms of $m_t-m$ at mid-rapidity is a consequence, a
special case of the more general universal scaling law of eq.~\ref{e:v2w},
predicted by the Buda-Lund model.

Furthermore, PHENIX found~\cite{Adare:2006ti,Afanasiev:2007tv}, that
$v_2/n_q$ scales in terms of $(m_t-m)/n_q$ for several types of particles
($\pi$, K, K$^0_s$, p, $\Lambda$, $\Xi$, deuteron, $\Phi$). This indicates
that the perfect fluid motion scales on the quark level, and the Buda-Lund
model scaling prediction for higher order flows, $v_{2n}=I_n(w)/I_0(w)$
(see ref.~\cite{Csanad:2003qa}) should also be applied on the quark level.

\subsection{Conclusions}
We have shown that the excitation function of the transverse momentum and
pseudorapidity dependence of the elliptic flow in Au+Au collisions is well
described with the formulas that are predicted by the Buda-Lund type of
hydrodynamical calculations~\cite{Csorgo:2001xm,Csanad:2003qa}. We have
provided a positive test for the validity of the perfect fluid picture of
soft particle production in Au+Au collisions at RHIC up to ~1-1.5\,GeV and
up to a pseudorapidity of $\eta_{\textrm{beam}}-0.5$.

We have also shown that the PHENIX
discovery~\cite{Adare:2006ti,Afanasiev:2007tv} of a scaling behavior of
$v_2$ vs. $m_t-m$ is a special case of the more general, rapidity
dependent universal scaling law of the Buda-Lund type of perfect fluid
hydrodynamical solutions.

The universal scaling of PHOBOS \veta and PHENIX and STAR \vpt,
expressed by eq.~(\ref{e:v2w}) and illustrated by Fig.~\ref{f:v2w}
provides a successful quantitative as well as qualitative test for
the appearance of a perfect fluid in Au+Au collisions at various
colliding energies at RHIC.

We have furthermore shown, that since PHENIX
found~\cite{Adare:2006ti,Afanasiev:2007tv}, that $v_2/n_q$ scales in terms
of $(m_t-m)/n_q$ for several types of particles, the Buda-Lund model
scaling prediction for higher order flows, $v_{2n}=I_n(w)/I_0(w)$ (see
ref.~\cite{Csanad:2003qa}) should also be applied on the quark level.

\begin{acknowledgments}
This research was supported by the NATO Collaborative Linkage Grant
PST.CLG.980086, by the Hungarian - US MTA OTKA NSF grant INT0089462 and by
the OTKA grants T038406, T049466, T047137. M. Csan\'ad wishes to thank
professor Roy Lacey for his kind hospitality at SUNY Stony Brook, and the
US-Hungarian Fulbright Commission for their spiritual and financial
support. We also would like to thank Arkadij Taranenko for his valuable
suggestions.
\end{acknowledgments}

\bibliographystyle{prlsty}
\bibliography{Master}

\end{document}